# Physics-informed motion registration of lung parenchyma across static CT images

Sunder Neelakantan[1], Tanmay Mukherjee[1], Emilio A. Mendiola[1], Kyle Myers[2], and Reza Avazmohammadi[1]

*Abstract* - Pulmonary hypertension (PH) can lead to significant vascular remodeling, resulting in altered pulmonary blood flow. Estimating the patient-specific contributions of each remodeling event is necessary to optimize and individualize clinical intervention strategies. In-silico modeling has emerged as a powerful tool to simulate pulmonary hemodynamics, and one of the primary requirements for robust in-silico modeling is an accurate representation of the pulmonary vasculature structure. Computed tomography (CT) imaging can be used to segment and reconstruct the proximal vasculature. However, contrast-enhanced imaging, such as CT pulmonary angiography, is required to obtain a comprehensive and high-fidelity view of the pulmonary vasculature. The clinical use of CT pulmonary angiography is limited by the complications associated with the injection of contrast agents. Machine learning (ML) approaches have emerged to effectively segment and reconstruct the pulmonary vasculature without the need for contrast-enhanced imaging. We have developed a method to create in-silico pulmonary angiogram phantoms with varying simulated contrast levels. The results indicated that adding simulated contrast can allow for successful segmentation of the pulmonary vasculature. We expect this method to assist with developing and training ML-based segmentation frameworks and aid in their validation, thereby improving the capability to segment and reconstruct pulmonary vasculature without using contrast-enhanced imaging.

*Clinical relevance* — This study can aid in the generation of synthetic data sets for the training and validation of ML-based segmentation and reconstruction tools, reducing the need

## I. INTRODUCTION

Pulmonary hypertension (PH), defined by elevated mean pulmonary arterial pressure (mPAP), affects approximately 1% of the global population, with prevalence rising to 10% in individuals over the age of 65 [1]. PH is associated with significant vascular remodeling, including stiffening and narrowing of blood vessels [2], as well as a reduction in the distal vasculature [3], [4]. These structural changes alter pulmonary hemodynamics [5], leading to impaired oxygenation and adverse alterations in cardiac function [6], [7]. The primary mechanisms of vascular remodeling have distinct hemodynamic consequences: vessel narrowing primarily increases resistance to blood flow, while arterial stiffening reduces vascular compliance and pulsatility. Considerable variability in the extent of these mechanisms has been observed, thus, investigating pulmonary hemodynamics on a patient-specific basis is critical to an improved understanding of the pathophysiological processes underlying the remodeling events that precede altered vascular hemodynamics and, subsequently, cardiac dysfunction. In-silico modeling has emerged as a powerful tool that can enable the investigation of in-vivo pulmonary hemodynamics in patients with PH [8]–[10].

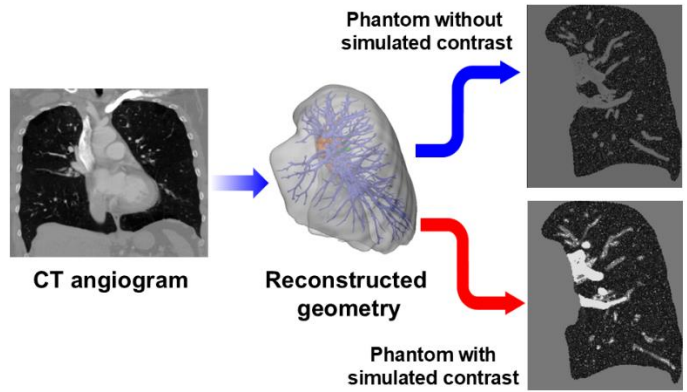

Fig. 1. Schematic of the reconstruction and in-silico phantom generation. CT angiogram images were taken from the Radiological Society of North America Pulmonary Embolism CT Dataset [11].

A primary requirement for rigorous in-silico modeling of pulmonary hemodynamics is an accurate representation of the structure of the pulmonary vasculature. Imaging modalities such as computed tomography (CT) and magnetic resonance imaging (MRI) are often be used to segment and reconstruct the proximal vasculature, i.e., up to 2-3 generations. However, contrast-enhanced imaging through CT pulmonary angiography is required to obtain the comprehensive structure of the pulmonary vasculature. To minimize the need for contrast, machine learning (ML) based segmentation and reconstruction have been introduced to isolate the structure of the airways and the pulmonary vasculature [12], [13]. However, while studies have shown the capacity of ML-based tools to isolate the structure of the pulmonary vasculature, the results cannot be easily validated without extensive contrast imaging data.

To overcome this limitation, we developed a method to create in-silico phantoms of PA angiograms, with the primary objective being to assist with the training and validation of ML-based segmentation tools. Pulmonary angiogram images were used to reconstruct the geometry of the lungs, the proximal pulmonary vasculature, the central airways, and the distal vessels as a finite element (FE) mesh (Fig. 1). The FE mesh of the airway elements was used to generate individual in-silico phantoms [14] of CT images. This method can enable the modulation of in-silico contrast, thus enabling the ML-based segmentation tools to be trained on various levels of simulated contrast. This work will facilitate the development of more robust and generalizable ML-based segmentation algorithms by providing a controlled platform for training and validation across a wide range of contrast

[1]Department of Biomedical Engineering, Texas A&M University, College Station, TX, USA
[2]Hagler Institute for Advanced Study, Texas A&M University, College Station, TX, USA

conditions, ultimately facilitating the patient-specific assessment of cardiopulmonary diseases such as PH.

## II. METHODOLOGY

*A. Pulmonary angiography*

CT pulmonary angiography images were obtained from the Radiological Society of North America Pulmonary Embolism CT Dataset [11]. One representative image was selected for segmentation and reconstruction (Fig. 1). 3D image convolution was performed following the equation:

$$[H * G](i,j,k) = \sum_{l=1}^{3}\sum_{m=1}^{3}\sum_{n=1}^{3} H(l,m,n)G(i-l,j-m,k-n), \quad (1)$$

where $G(i, j, k)$ and $H(l, m, n)$ are the 3D image and convolution kernel, respectively. In Eq. 1, $i, j, k$ represent height, width, and depth of the image, respectively. The 3D convolution was used to perform a low-pass filter to reduce background noise and smooth the image, for which the 3D kernel is given by:

$$H_1(i,j,k) = \frac{1}{36}\begin{bmatrix} 1 & 1 & 1 \\ 1 & 2 & 1 \\ 1 & 1 & 1 \end{bmatrix}, \quad k = 1,3, \quad (2)$$

$$H_1(i,j,k) = \frac{1}{36}\begin{bmatrix} 1 & 2 & 1 \\ 2 & 4 & 2 \\ 1 & 2 & 1 \end{bmatrix}, \quad k = 2. \quad (3)$$

*B. Segmentation and reconstruction*

The representative CT image was segmented and reconstructed using the Materialize Mimics Innovation Suite. The segmented geometry was smoothed to remove the airways and the pulmonary vasculature. Next, the central airways were isolated by segmenting the low-intensity regions connected to the trachea. The proximal pulmonary arteries and veins were segmented by isolating solid high-intensity regions connected to the heart. To isolate the distal airways and vessels, 3D convolution (Eq. 1) was performed using a 3D edge detection kernel defined by:

$$H_2(i,j,k) = \begin{bmatrix} 0 & 0 & 0 \\ 0 & -1 & 0 \\ 0 & 0 & 0 \end{bmatrix}, \quad k = 1,3, \quad (4)$$

$$H_2(i,j,k) = \begin{bmatrix} 0 & -1 & 0 \\ -1 & 6 & -1 \\ 0 & -1 & 0 \end{bmatrix}, \quad k = 2. \quad (5)$$

For reconstruction, the central airways were subtracted from the lungs. The geometry of the lungs with the central airways removed was used to create a volumetric mesh consisting of tetrahedral elements with a target edge length of 5 mm. In addition, the proximal vasculature and distal vessels (a combination of airways and vasculature) were used to create volumetric meshes, with a target edge length of 2 mm.

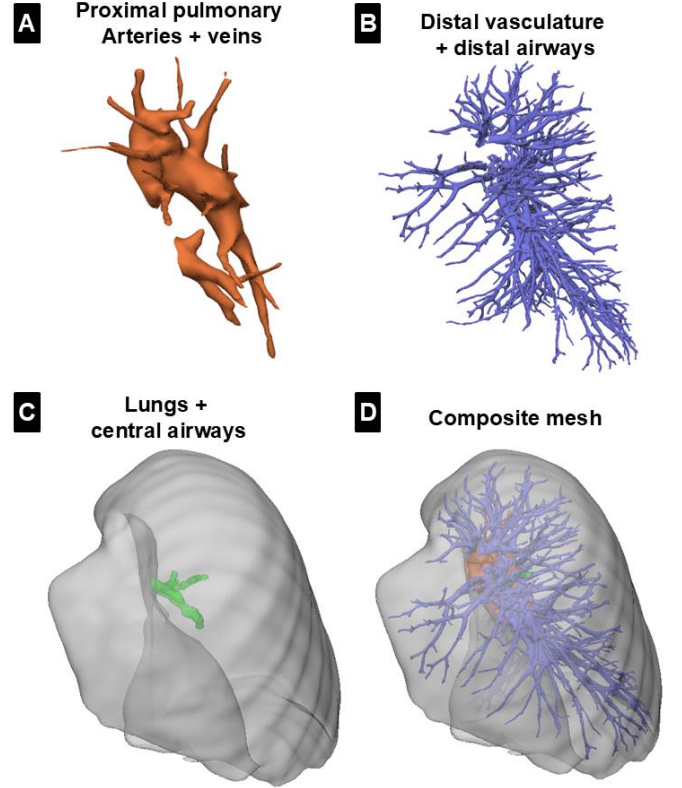

Fig. 2. Reconstructed geometry from CT angiogram. (A) Proximal pulmonary arteries and veins. (B) Distal pulmonary vasculature and airways. (C) The left lung with the central airways removed. (D) The final composite mesh used to generate in-silico phantoms.

*C. In-silico phantom generation*

The lung geometry was used to generate an in-silico phantom based on our previous work [14]. Briefly, the FE lung mesh nodes were mapped onto a uniform grid using the visualization toolkit (VTK) with predefined parameters. The intensity values of the grid points, which correspond to voxels, were controlled by a proximity threshold, i.e., the number of node points of the mesh within a sphere of a chosen radius centered at a point in the uniform grid.

In addition, the proximal and distal vessels were used to generate two additional in-silico phantoms. The voxel size for these phantoms was set to 1 mm, and the proximity threshold was set to 2 mm. These phantoms were added together, and the intensity of the regions containing tissue and blood was normalized to the same value to simulate the absence of contrast. Next, the phantoms were added, with the proximal vessel phantom weighted higher to simulate the presence of contrast in the pulmonary vasculature.

*D. Comparison between mesh reconstructed from image and from phantom*

The in-silico phantom generated with the simulated contrast was used to segment and reconstruct the proximal pulmonary vasculature. Next, the mesh generated through the phantom was compared against the mesh generated from the original image

through the use of the Hausdorff distance error metric ($d_H(X, Y)$) [15], which was given by

$$d_H(X,Y) = max\left\{\sup_{x \in X}\inf_{y \in Y} d(x,y), \sup_{y \in Y}\inf_{x \in X} d(x,y)\right\}, \quad (6)$$

where inf and sup are the infimum and supremum functions, respectively. x and y are nodes belonging to the node sets X and Y respectively. d(x, y) is the distance between the points x and y. Briefly, for a node x in node set (mesh) X, the minimum distance to node set Y is estimated. The Hausdorff distance is the maximum value of such distances. The final Hausdorff distance was normalized against the characteristic length of the vasculature mesh (140 mm).

## III. RESULTS

The lung geometry, central airways, and proximal pulmonary arteries and veins were successfully segmented from the CT angiogram through the appropriate filtering (Figs. 2A, C). The accuracy of the proximal pulmonary vasculature was improved by using a sharpening filter. The distal vasculature and airways could not be segmented accurately without filtering and required an edge detection filter to successfully segment and reconstruct (Fig. 2B).

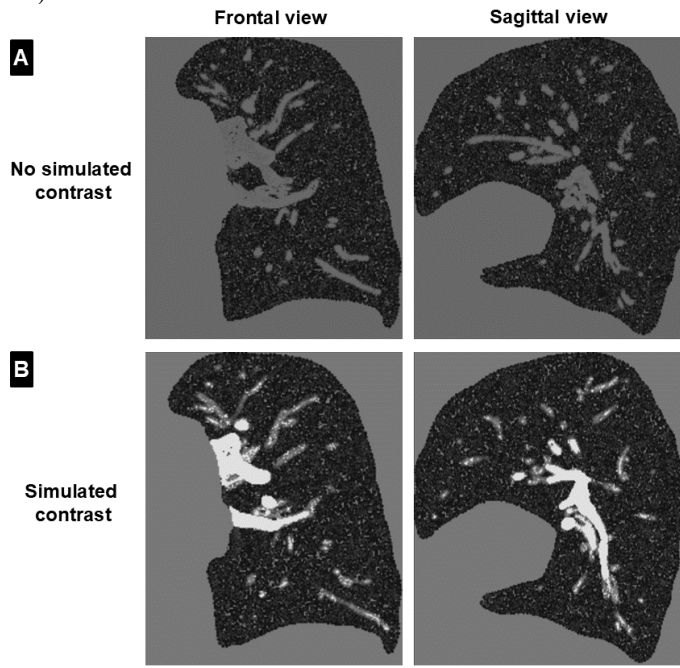

Fig. 3. Fig. 3. (A) In-silico CT phantom without the effect of simulated contrast. (B) In-silico CT phantom with simulated contrast effect.

After the segmentation and reconstruction, the in-silico phantom of the individual components was generated. This created three effective "layers" - the lungs and central airways, the proximal pulmonary arteries and veins, and the distal components (vasculature and airways). Using the individual components, a composite phantom was generated with the intensity normalized, such that the vasculature could not be separated from the airways (Fig. 3A). Next, the intensity of the proximal vasculature layer was increased to simulate the effect of contrast injection (Fig. 3B). The increased intensity of the proximal vasculature improves the ability to visualize and segment the pulmonary vasculature.

Comparing the geometries of the pulmonary vasculature generated from the original CT angiogram and the in-silico phantom, the major difference observed was the lack of sharp details in the smaller vessels (Fig. 4). The rounded features can be attributed to the voxel size being too large to capture the features accurately, suggesting a need to reduce the voxel size. In addition, the phantom generation methods require further improvement to capture similar levels of detail as the actual images, and this will be addressed in future studies. The Hausdorff distance was found to be 3.8 mm, which, when normalized by the characteristic length of the vasculature of 140 mm, yields a percent error of 2.14%. Improving the phantom resolution can minimize the overlap between different components and further reduce the error metric.

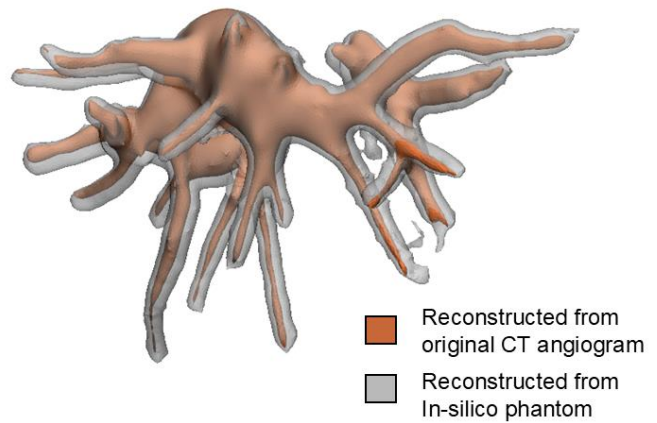

Fig. 4. Comparison between the pulmonary vasculature geometry reconstructed from the original image and from the in-silico phantom. Orange - reconstructed geometry from original CT angiogram, gray – reconstructed geometry from in-silico phantom.

## IV. DISCUSSION

In this study, we have presented a method to generate synthetic CT images of the lungs comprising parenchyma, airways, and pulmonary vasculature. In addition, multiple levels of contrast in the pulmonary vasculature can be simulated to generate in-silico phantoms resembling conventional CT images and CT angiograms. Using the method presented herein, ML-based segmentation tools can be trained to reconstruct the pulmonary vasculature without the need for contrast-enhanced imaging. An estimated 70 million contrast-enhanced scans were performed in 2019, with a risk of acute allergic reactions of 0.6% [16]. Thus, the elimination of the need for injected contrast agents could prevent approximately 400,000 yearly acute allergic reactions. In addition, ML tools can be trained on the same phantom image at multiple simulated contrast levels, thus improving accuracy, and the presence of the "ground-truth" FE mesh serves as a benchmark for model validation. The ability to accurately segment the pulmonary vasculature with and without contrast can enhance the translational potential of tools that utilize the geometry of the pulmonary vasculature to model and predict remodeling in cardiopulmonary diseases, such as PH.

Despite the potential benefits of the presented method, its current fidelity leads to a few key limitations. The primary limitation is the need for initial manual segmentation and reconstruction. Future studies will involve developing a method to automate the segmentation and reconstruction process. Once reconstructed, the generation of the in-silico phantoms at multiple simulated contrast levels can be performed automatically with the current method. In addition, generating in-silico phantoms is time-consuming, but this can be significantly sped up by performing this computation on graphics processing units (GPUs), where many voxels can be estimated in parallel. The in-silico phantom was generated at a voxel resolution of 1 mm, which is comparable to high-resolution CT imaging in the clinic. Increasing the resolution leads to additional computational costs, as well as surpassing the resolution of images acquired in clinical settings. This could potentially lead to poor training accuracy for ML models as the phantoms and actual images are at different resolutions. Future studies will be focused on improving the method used to generate in-silico phantoms to better recreate the actual CT imaging process. In addition, future work will focus on using deep learning-based filtering and segmentation methods to improve the accuracy of the reconstructed geometries.

## V. CONCLUSIONS

In this study, we have presented a method to generate synthetic CT images of the lungs that can simulate varying levels of contrast. We expect this method to aid in generating synthetic data sets for training ML-based segmentation and reconstruction tools and validating such tools. In addition, the geometry of the pulmonary vasculature reconstructed from the images can be used for computational methods such as virtual catheterization.

## COMPLIANCE WITH ETHICAL STANDARDS

This research study was conducted retrospectively using human subject data made available in open access by Radiological Society of North America Pulmonary Embolism CT Dataset [11]. Ethical approval was not required, as confirmed by the license attached to the open-access data.

## ACKNOWLEDGEMENT

This work was supported by the National Institutes of Health (R56HL172052) to R.A. and the American Heart Association (24PRE1240097) to T.M.